\documentstyle[12pt,aaspp4]{article}

\begin{document}

\title{Imaging of z$\sim$2 QSO host galaxies with the Hubble Space Telescope
\footnote{Based on observations with the NASA/ESA Hubble Space telescope,
obtained at the Space Telescope Science Institute, which is operated by
the Association of Universities for Research in Astronomy (AURA) Inc,
under NASA contract NAS5-26555.}}

\author{J.B. Hutchings, D. Frenette}
 \affil{Herzberg Insitute of Astrophysics,
National Research Council of Canada,\\ Victoria, B.C. V8X 4M6, Canada\\
john.hutchings@nrc.ca}

\author{R.Hanisch, J.Mo\footnote{Computer Sciences Corporation}}
\affil{Space Telescope Science Institute,\\3700 San Martin Drive, Baltimore,
MD21218\\hanisch@stsci.edu,jinger@stsci.edu}

\author{P.J. Dumont, D.C. Redding}
\affil{Jet Propulsion Laboratory,\\4800 Oak Grove Drive, M/S 306-438, Pasadena,
CA91109\\david.c.redding@jpl.nasa.gov,philip.j.dumont@jpl.nasa.gov}

\author{S.G.Neff}
\affil{Code 681, NASA Goddard Space Flight Center, Greenbelt, MD20771\\neff@stars.gsfc.nasa.gov}

\begin{abstract}
    We report on deep imaging in 2 filters with the PC2 camera of HST,
of five QSOs at redshift $\sim$2, with a range of optical and radio 
luminosity. The observations included a suite of PSF observations 
which were used to construct new PSF models, described elsewhere
by Dumont et al (2001). The new PSF models were used to remove the
QSO nucleus from the images. We find that the host galaxies have resolved 
flux of order 10\% of the QSO nuclei, and are generally luminous and blue,
indicating active star-formation.
While most have clearly irregular morphologies, the bulk of the flux can
be modelled approximately by an r$^{1/4}$ law. However, all host galaxies also
have an additional approximately exponential luminosity profile beyond a
radius about 0.8 arcsec, as also seen in ground-based data with larger
telescopes. The QSOs all have a number of nearby faint blue companions 
which may be young galaxies at the QSO redshift. We discuss implications
for evolution of the host galaxies, their spheroidal populations,
and central black holes. 

\end{abstract}


\section{Introduction}

   There have been several investigations of QSO host galaxies at redshift
2 and higher, using 4m class telescopes (with and without adaptive optics
image correction), and with the Hubble Space Telescope. Many of these
are reviewed by Hutchings (2001a). Heckman et al (1992) and Lehnert et al (1992)
were the first to discover that high redshift QSO hosts were both large and
luminous (particularly of radio-loud QSOs), and unlike galaxies in the
present-day universe. However, there is a range of host galaxy sizes and
luminosities, and it is not easy to study their detailed morphology and
colour because of size, redshift dimming, and the presence of the bright
central QSO nuclei. In general, however, the QSO hosts become more luminous
and compact with increasing redshift, and are found in regions of enhanced
faint galaxy counts. There have been claims regarding the host morphology,
based on fits to luminosity profiles of the hosts after removal of the central
source (e.g. Kukula et al 2001, McLure et al 1999). 

   The host galaxies of higher redshift QSOs are of interest for several 
reasons. QSOs show strong cosmic evolution in their population, luminosity,
and radio morphology, and we wish to see how the triggering and fuelling 
of the QSO episodes causes this evolution. Seen over large redshift ranges,
the QSO hosts must undergo significant evolution as galaxies, and offer
an opportunity to study evolution of galaxy star-formation, stellar 
populations, merging, and morphology changes in addition to the nuclear
activity. It is clear from local galaxy studies that the central black hole
mass is closely related to the spheroidal stellar population, so that we may
seek to understand how this population forms and evolves, as well as how the
central black hole forms and grows with time. We may also measure the central
mass independently by the width of the broad emission lines in the QSO
nucleus, providing another link to galaxy and black hole evolution.

   The results reported here are from a program proposed in 1995, when these
studies were new, with a small total sample. As table 1 shows, the 
observations were not performed until several years later, and other studies
had been undertaken. The sample chosen was intended
to cover a range of luminosities in both optical and radio, with some
complementary ground-based coverage, and restricted to a small redshift
range close to 2. We describe below further details of the observational
procedure. We are particularly concerned with accurate modelling and
removal of the central point spread function (PSF), on which many of the
above conclusions rest.

\section{Observations}

    The observations were carried out as a delayed cycle 6 HST program, and 
are listed in Table 1. The QSOs were centred in the PC CCD of the WFPC2,
and exposed
for about 5000 sec in each of the F606W and F702W filters, using several
readouts at each filter, to enable cosmic ray removal and to avoid saturation
in the nuclear pixels. The rest wavelengths sampled with the filters
are in the range 1800 to 2300\AA, so are sensitive to young stellar
populations. In this paper we refer to the filters and magnitudes as
R and I, since the conversions are within the measuring errors of our
results.

   The images from each QSO were combined and
cosmic rays removed in the usual way. No unusual problems were encountered
and there were few saturated pixels to deal with. These occurred in the image
centres where we have no resolved information. The pixels were replaced with
values from scaled PSF images, and were given zero weight in the PSF removal
process described below. Obvious bad pixels and cosmic ray pixels were
interpolated across in the QSO images.

 In addition to the
QSOs, a set of PSF observations was made of the star Feige~23, chosen
to have an SED in the optical similar to the redshifted QSOs in the 
program. The PSF
observations involved a range of exposure times and dither positions around
the centre of the PC field, and were used to generate a model for the
PSF for the QSOs.

The PSF observations consisted of four
sequences of exposures of a star in each of the two filters. These four
sequences were at four field positions: the center and the vertices of 
a triangle surrounding the nominal PC chip position for the QSOs. The 
triangle was $\sim$3 arcsec on a side, to match the nominal positioning 
accuracy of targets on the WFPC2 focal plane, oriented at an angle with
respect to the CCD rows and columns, to provide both minimum overlap of PSF
diffraction features and maximum prescription retrieval fitting leverage in
determining alignment parameters. Each exposure sequence contained pairs of
exposures with durations of 0.23 sec, 4 sec, and 100 sec. The sequences were
chosen to avoid pixels affected by the residual image from previous saturated
exposures. 

The paired exposures were used to both minimize the effect of 
noise in the prescription
retrievals and to identify cosmic ray events. The shorter exposures were
designed to properly expose the PSF core and enable accurate registration 
of the PSF with the QSO cores. The 0.23 seconds is the minimum
WFPC2 exposure for which the PSF characteristics are unaffected by the
shutter flight time. This range of exposure times provided measurements of 
PSF features over the dynamic range of interest in the the science 
observations; PSF features
that are factors of 10$^4$ and 10$^5$ weaker than the core were measured 
at S/N approximately 35 and 8.5 respectively.

\section{Point spread function removal}

   The PSF modelling work required for this project was carried out as
described in detail by Dumont et al. (2001). We also constructed empirical 
PSFs from the Feige 23 observations, and Tiny Tim PSF models for the QSO
locations in the PC (Krist 1995).  However, our analysis was done entirely 
with the Dumont et al. model PSFs, as these resulted in 
significantly cleaner removal of PSF structures.

  The Dumont et al PSFs use a hybrid approach to modelling
the HST optical system and WFPC2 PC camera, by combining prescription 
retrieval to solve for the primary optical components, and phase
retrieval to compute a "mirror map" giving deviations in the HST primary
mirror from the nominal prescription.  The initial prescription (low-order
Zernike coefficients) and mirror map were estimated using a series of out of
focus images (generated by relocating the secondary mirror) obtained from
the HST archive.  The prescription was refined by refitting parameters known
to vary (secondary mirror position) and known to be dependent on the
particular observation (object position in field of view, object intensity,
background) and taking into account the object's spectral type and filter
bandpass.  

  The actual PSF models used were ultimately chosen by matching the inner
structure and main diffraction spikes in the PSF models with those in 
the QSO images, as described below. The models were computed with a pixel 
size eight times smaller than the actual observations, allowing us to 
make sub-pixel offsets without introducing sampling errors.  
The final parameters that were
adjusted in the PSF models were a) the SED and emission line flux of the
program QSOs within each filter bandpass, based on published spectra or the
broad-band colours and a `standard' QSO spectrum, and b) the optical offset
of the secondary mirror.

   It was found that the flux distribution within the filter bandpass is
not a critical PSF issue, with very small changes betwen a flat SED and one
that reflects the continuum and emission lines within the bandpasses.
The principal differences between the filter bandpasses are the scale
changes in the PSF structure that are caused by the mean wavelength of 
the bandpass. 

  On the other hand, the optical offset of the HST secondary mirror does 
make a major difference to the detailed intensities of the inner PSF 
structure and the relative intensities of the main diffraction spikes 
(see Figure 1). The PSF has three distinct `zones': the inner very bright 
few pixels, a ring of 12 complex knots and `streamers' out to radii of 
about 0.3 arcsec (at our filter wavelengths), and four diffraction spikes
that are seen to large radii. In our PSF-fitting, we found the inner
few pixels cannot be fit well, partly due to saturation effects in the
data, and anyway contain no resolved information. Thus, the criteria 
for choosing the PSF model are the relative brightnesses of the
ring of knots and the long spikes. A best model was selected from a grid 
of secondary mirror offsets for each QSO and filter. The model grids
were done with a coarse grid and a finer grid near the best-looking
value. The PSF model selection was done by inspection and then by 
minimising the residual structure in the PSF-subtracted QSO images. 
It was found that the same model offsets
(close to 1mm in both x and z) were selected for all QSOs and both 
filters, independently by the first two authors. These were significantly 
better than Tiny Tim or a simple mean PSF from our grid of observations,
in matching the detailed structure seen in the QSO images.

  After selecting the best-looking PSF model, it was block-averaged by a 
factor 8, 
using a grid of 8 x 8 single (fine) pixel offsets. This enabled us to match 
the pixel sampling of the QSO, which was judged by the sampling of the four
triple-stripe diffraction spikes. These spikes are about one PC pixel wide.
This was the PSF used with matched sampling to the data, for subtraction
from the QSO images. We found that the subtraction results do not change
rapidly within 1mm offset steps, so that selection of the
`best' model was not critical within a range of about 0.3mm in the offsets.

  The PC-pixel sampled PSF was then scaled and subtracted from the QSO image.
In all cases a grid of scale factors (and pre-block-averaging PSF shifts) 
was used to decide 
which was the best PSF subtraction. In choosing the best subtraction, the
subtracted image was modelled for its fit to exponential and
de Vaucouleurs law azimuthally averaged profiles. The subtracted images
were characterised by using the `ellipse' task in IRAF, centred on the
nucleus position. The fits were judged by minimising profile differences
from the two types of model - i.e. linearity in log(signal) against radius or 
radius$^{1/4}$. We gave zero weight to the innermost 4 pixel radius
(about 0.65 in R$^{1/4}$ in Figure 3),
which lie inside the resolution of the images and where there are saturated
pixels; and a weight inversely proportional to the error bars calculated from 
the image by the ellipse task.   In the diagrams and plots, we have 
interpolated across the low pixels in the inner 4 pixel radius, 
resulting in
an apparent excess of unresolved flux in the central pixels. We stress that
there is no information in these pixels and they are not taken into account
in the measured quantities. Best fit models are extrapolated into the central
few pixels in deriving the quoted host galaxy fluxes.

   We also inspected the 2-dimensional subtracted images for residual PSF
structure. This was already minimised in morphology by the PSF model selection,
but visal inspection is sensitive to under- or over-subtraction that might 
not show up in azimuthally averaged profiles. The final adopted subtractions
were a mean of the profile fits and image structure inspection values for
scaling, but these were always the same within a few percent in the final
resolved flux estimates. 

   Finally, the sensitivity to the PSF models, positions, and scaling were
tested by finding best fits over a range of each of these parameters. The
PSF-subtracted fluxes were robust to within 10\% - i.e. for the range
of fits that are acceptable by the above criteria. The host galaxy structures
outside radii of 0.5 arcsec are not sensitive to any of the fit
parameters explored, while structures and total signal level
inside this depended on the relative nuclear to host flux ratio, as discussed
individually below. 

  Figure 2 shows an example of PSF-subtraction from a PSF-star image. The
diagram illustrates that the central region is not properly modelled (by
choice, as noted above), and that no detectable flux is seen in the
subtracted image beyond radius 0.35". This PSF fit is not as good as those
for the QSOs, as the short PSF-star exposures suffer from telescope breathing
and worse read noise than the longer QSO integrations, and are from different
places on the detector, so we did not attempt very detailed modelling for 
the individual PSF observations. Neverthless, we note that all the QSOs
showed more extended and significant resolved structure than the star image
in Figure 2.

   For each PSF-subtracted QSO image, we calculated (and tabulate) the 
total flux from 
the best-fit model, which interpolates across the central pixels as shown
in the diagrams. The values agree well with the summed image signal
within the radius limits of good fit. \it In all cases we found much better 
fits for the inner host galaxies with a de Vaucouleurs model, but 
in most cases this did not
fit well beyond a radius of about 0.8 to 0.9 arcsec. \rm Beyond this, we
applied an exponential model fit: a different de Vaucouleurs model fit as
well, but gave about the same total flux, and we consider the outer
galaxy more likely to be an exponential anyway. We also note that the
images show that the host galaxies are not at all symmetric ellipses,
so that \it we are using the de Vaucouleurs models as flux estimators
rather than morphology discriminators. \rm

    We describe the results for each QSO in the sections
below. Table 2 summarizes the adopted best-fit subtractions and the
measured or model quantities from them. The diagrams show subtracted 
images and luminosity profiles for all the objects. The plots have limiting
surface brightness of about 27 mag/arcsec$^2$. With smoothing it is
possible to go about a magnitude fainter at radii about 3 arcsec, but
the signal is poor and very sensitive to the sky level uncertainty.
Note that we have used cosmological parameters H$_0$=100 and q$_0$=0.5
in quoting absolute magnitudes, with no K-correction applied.

 We recorded the fluxes and positions of all galaxies measurable in both 
filters in the PC field
surrounding the QSO. As all exposures and QSO redshifts are similar, this
is a reasonably good comparison of the galaxy environments of the sample
QSOs, in terms of richness and galaxy colours, down to F602W magnitude 
$\sim$26. However, this part of the work is not a rigorous or flux-limited
investigation of the QSO companions, as we were unable to measure fainter
or LSB objects that were visible in some fields.

\section{Notes on individual QSOs}

\subsection{0033+098}

   The QSO is radio-loud (460mJy at 5 GHz: Pauliny-Toth et al 1972), 
but with no known radio 
structure. The QSO was fainter than the catalogued V=17.5 (Hewitt and 
Burbidge 1993) suggests, with R=18.9 and I=18.6. 
We have measured several nearby faint galaxies, and most have similar
colours (see Table 2). 

   Figures 3 and 4 show the images and luminosity profiles for the QSO.
The best-fit de Vaucouleurs model is shown, as discussed in the previous
section. In fact, the QSO host galaxy appears to be asymmetrical and to 
contain knots, so the assumption of a smooth de Vaucouleurs profile is not
necessarily the best way to measure the host galaxy. The `ellipse' IRAF task
used can be used to fit an ellipse to successive mean radii with a free
centroid, so that the effect of departures from azimuthal symmetry are 
minimsed, while allowing us to quantify deviations from central symmetry 
by the migration of successive ellipse centroids.

   In both F606W and F702W images, the de Vaucouleurs fit was good out to
a radius of about 1". Beyond that, there is excess light that may be
approximated better by an exponential disk (or different de Vaucouleurs
model), out to about 2" radius. Beyond that, 
the signal is too  weak to measure. This illustrates a generic point
well: HST has good resolution and signal in the radius range 0.2" to 1.0".
Ground-based observations with larger telescopes reveal flux extending to
several arcsec in many such objects. Thus, HST is measuring the central
spheroidal-like population of the host galaxy, but may fail to detect
a more extended disk, arms, or halo. Our deep HST images just cover the
transition between these regions.

  The host galaxy does have asymmetrical structure that is seen (Figure 1)
as an extension to the top of the image, plus a separate knot to the left
side. The fitted ellipses have centroids that move systematically
by several pixels from the nucleus to the outer contours, reflecting this.
Note too that the visible structures lie within the radius fit by the
de Vaucouleurs model, so that the `bulge' model fit does not imply the 
presence of a smooth bulge in the conventional sense.
These structures are faint, but the overall resolved flux from the host
galaxy indicates a high luminosity, which is several magnitudes brighter
than any of the nearby galaxies, and corresponds to absolute magnitude
M$_R\sim$-25. This is similar to measurements of other radio-loud QSO
hosts at redshift near 2.

   Table 2 summarizes the measurements of this and the other QSO fields.
Other galaxies seen in the PC frame are faint and compact, and are 
generally blue, with two exceptions at R-I = 1.6. In all the QSO fields
we have not counted galaxies redder than R-I=1.0 in Table 2. The other fields
have only one such red galaxy each, so they form a minor part of the
companion totals. The flux from the extended structure to the N of
the QSO, measured on its own, is also blue, with R-I close to 0.0. 
The `companions' range in luminosity from our Galaxy to the LMC if at 
the QSO redshift. They are all blue enough to be star-forming at the 
QSO redshift or higher. Sizes range from small to moderate (0.3 to 1.5 arcsec) 
to the brightness level detected. The space density at the QSO redshift is high.

\subsection{0225$-$014}

    This is a radio-loud QSO (4C 01.11: 300mJy at 2.7GHz,150mJy at 5GHz:
see e.g. Barthel et al 1990). 
It has C IV absorption at z=2.0. It is a triple radio source (17" long)
with a 35$^o$ bend, and an overall spectral index fairly steep at 1.1. 
Table 2 shows the measured quantities for the QSO and associated objects.
The QSO host galaxy is resolved and has quite asymmetric structure. The
PSF removal works well outside the central 6-7 pixel diameter, and is
shown in Figures 3 and 4.

   The azimuthally averaged profile from fitting ellipses to the PSF-subtracted
QSO fits a spheroidal model fairly well, but there are significant bumps and
dips in the profile. The outer parts (beyond radius 0.8") in the F606W
image can be fit to an
exponential profile, and the spheroid model does not fit in the F606W
filter. Thus, there
may be a faint halo or `disk' that we are just detecting. In the
F702W image, the arm is less conspicuous and the disk component is not
indicated: a spheroidal model fits to the radius limit of the image. 
We stress that these model
fits are useful in estimating the resolved total luminosity but do not
indicate the presence of normal bulge or disk morphology. The resolved
flux is one-sided and contains a curved bright arm that extends almost 1"
from the nucleus. The centroids of the ellipses move some 0.2" from the
nucleus to mean radii of 0.7" as they include this arm.

   The radio structure has only ~1.4" resolution but shows an extended core
(jet?) whose direction is along the brightest HST diffraction spike residual.
Thus, our host galaxy detection is poorest along this line. The curved arm
of the host galaxy appears to originate at this angle. The radio lobe peaks 
lie within the surrounding group. The NW lobe is along the jet direction
and is further from the nucleus by about a factor 2. 
It does not correspond with any detected
optical source. The SE lobe is extended perpendicular to the radius to the
QSO and lies near the edge of the group of galaxies, close to
(but not on) a diffuse large faint galaxy. 

     The QSO lies in a field of faint companions. 26 are seen in the PC chip
(although only 16 were measured reliably in both filters),
which is about double that seen in others in the sample. The companions
are grouped within the PC field, with the QSO off-centre. 
Only one or two have colours that
suggest they are foreground - the rest are very blue and thus may be
star-forming at high redshift. The galaxies range from very compact to
small and nucleated, many with very asymmetrical structure. Unusually, there
are also several (7) which are large, LSB, and not nucleated at all. The
largest is blue and quite luminous.

    The QSO appears to be part of a group of young galaxies or protogalaxies
and to be in a host galaxy that is large and irregular, with low surface
brightness features. In view of the absorber seen at the slightly lower
redshift of 2.0, some (or all?) of the resolved structure and of the 
companion galaxies may lie  in this group at a few thousand km/s lower 
redshift than the QSO.

\subsection{0820+296}

   This is a radio-loud QSO that has been imaged with the CFHT in earlier
work (see Hutchings 1995 a,b and references therein). 
The ground-based visible imaging at 0.6" FWHM resolution indicated that
the QSO is extended azimuthally at radii 1.5" to 4".
The QSO was also found to have a high count of nearby galaxies at projected 
distances 20-30", with magnitudes
complete to about R=24. Many of them are blue, and were
discussed in terms of star-forming companions at redshifts similar to the QSO.
The QSO spectrum contains absorptions at z$\sim$2.05, so the galaxies may be
part of that absorbing group. 

    The resolved light in the F606W filter has a ring-like structure, somewhat
offset and with brightness that changes around the ring. The F702W image 
shows this less clearly but has more resolved light overall. Figure 2 shows
the luminosity plots and the adopted spheroidal models that fit them best
(but with significant irregularity in the profile).
The F702W image also has a halo or outer disk beyond a radius of 0.9".
This is seen less significantly and with less light in the F606W image.
It is possible that this galaxy has a ring of new star-formation plus an
older population, indicating an event that triggered the QSO episode. 

    The WFPC2 images show fainter galaxies than the ground-based
(about 1 magnitude, where they are compact). This changes the selection 
criteria for galaxy counts from those in the CFHT study. There are 12
measured in both colours on the PC chip, plus one star which is the brightest
and reddest object. One galaxy is red (and bright) and is likely to be
foreground. The rest are all blue and thus possibly star-forming at high
redshift, as noted in the CFHT study.

\subsection{1338+277}

    The QSO is radio-quiet, discovered in the Crampton et al (1995) survey. 
They note it is
extended. The QSO was imaged with the CFHT (Hutchings 1995 a,b) with 0.8"
FWHM image quality. The azimuthally averaged profile is resolved at radii
1-3", and a knot 1.2" to the North is noted. There appears to be some excess 
of galaxies to R=24 around it to radii approx 30". 

   The new data show the QSO is faint (about 21.5 magnitude) and the HST 
images have little
visible PSF structure. The QSO is non-circular down to very small radii
(0.2"), and has faint flux with elliptical contours out to some 0.6".
The images also clearly show a jet-like extension at some 60$^o$ to
the inner structure, extending some 1.4" and ending at a brighter knot
(see Figure 3).
The knot is more compact in F606W and there is an apparent gap between the
`jet' and knot.  

   The elliptical inner host extension is blue and the inside part of 
the jet is red. The QSO has R magnitude 1.6 fainter than the catalogued 
V value of 20. The jet is 26.7m and the blue region comparable. 
The `bulge' component is as shown by the plotted models and the `disk' 
is an additional exponential beyond radius 0.7".
These fits produce reliable total flux values, but are clearly not good 
fits to the morphology.

   The PC chip contains 12 companion objects, measured in both filters. The QSO
is the brightest object in the PC chip. The companions are small and faint, 
with 2 exceptions, but all could be galaxies at the QSO redshift. 
All but one (which is also one of the two larger ones) are blue objects, 
so consistent with their being true companions.
The area of sky covered by this group is within the PC field - about 
180 Kpc on a side - so that
this is a high concentration. The galaxies have luminosities like the LMC
or M33 if at the QSO redshift and with low k-correction.

\subsection{2244$-$010}

      This is an LBQS object (Foltz et al 1989), so has a luminous 
nucleus, and is the brightest
in our sample by over a magnitude. It has associated
absorption in the C IV and L alpha profiles (not BAL, but narrower). Little 
else seems to be known, and no radio detection is reported. 

    The HST images show very little structure and the bright nucleus
makes imperfections in the PSF modelling and subtraction more significant.
Our best PSF subtractions suggest
a small compact knot about 2.5arcsec to the West, but otherwise no
obvious structure beyond diffuse extended light through the PSF features.
Figure 4 shows that there is no
good fit to a spheroidal model, especially at the shorter wavelengths. 
An exponential works moderately well beyond radii of 0.8 arcsec.
The host galaxy light is redder
than the nucleus, and similar to other faint galaxies in the field.

   There are several faint and compact companions, all blue enough to be
at the QSO redshift. One galaxy is red (and larger) and is presumably
foreground. One other largish (chain?) galaxy is blue and quite luminous
if at the QSO redshift. The spatial distribution is even over the PC chip,
and thus not obviously clustered around the QSO. The companions range from 
very compact to diffuse with no nucleus (but still small).

\section{Discussion}

   Our new PSF models have been instrumental in achieving improved
resolution and modelling of the inner parts of high redshift QSO hosts.
However, there are still PSF artifacts that are not properly removed, most
obviously in the brightest of our targets. It is not straightforward to
quantify our errors. We have covered a wide range of PSF shifts and scale 
factors, as well as PSF models in this work, and consider that
the resolved flux values quoted are robust to a level of 10\% in all cases.

   Table 2 compares the principal results for the sample. Since the
QSOs are all at similar redshift and have similar exposures, the spatial 
scales and apparent colours may be compared directly, as well as properties
of companions. In the Table 2 summary, we have removed all companion galaxies 
that are redder than R-I=1 mag, supposing that they may be foreground objects.
There are few of these and the average numbers are little affected by their
removal. 

   In all cases, the bulk of the resolved light is best fit by a spheroidal 
type of profile, but in all cases this requires an extra component
which we have modelled with an exponential, beyond radii
of about 0.8" (corresponding to about 4Kpc in the adopted cosmology). 
This statement is robust against uncertainties in the
sky level, which affects only the outermost 2 or 3 points in our plots.
The exponential light is faint and only its innermost parts
are detected in our data. In ground-based observations with larger telescopes
(e.g. Hutchings 1995a), the faint outer light is often resolved out to several
arcsec, further reinforcing the model of a two-part light distribution for 
these galaxies, which we can `fit' with spheroidal and disk models. However, 
the morphology is in all cases irregular within the radius of bulge model 
fit, and indeed some of the luminosity profiles themselves are irregular, 
so we do not simply interpret the azimuthally averaged profiles in terms 
of bulge and disk components seen in regular galaxies in the local universe,
and do not conclude that the host galaxies are all `elliptical'.
We also note that less deep HST imaging would fail to detect any of the outer
`disk' light. As noted by Hutchings (2001a,b) we should thus be cautious of
claims that HST data reveal that high redshift QSO hosts are all `elliptical'.

   The resolved flux, however, is quite well measured by these models,
and is 2-3 magnitudes more luminous than L* in the local universe. 
The colours are blue, and with rest wavelengths in the NUV, indicate
that the light is dominated by massive young stars. 
Note that the k-corrections for a young population at this
redshift are negative, but the amount depends on the age (and reddening)
of the young stellar population present. Thus, comparison with present
epoch standard galaxy luminosities is not very meaningful. If free of
dust, and evolving passively without further merging events, the QSO host
galaxies would become present day galaxies of about L* luminosity.

There is no difference
between radio-loud and radio-quiet QSO hosts in our small sample, but
the two radio-quiet QSOs are the brightest and faintest of the group, and
their discovery techniques (LBQS and faint optical search) certainly bias
the comparison. The QSO R and I magnitudes of four of the group are 
sufficiently fainter than their catalogued V magnitudes, that they must have
varied. Overall, the resolved light lies close to 10\% of the QSO for all
objects, averaging the V, R, and I values.

  The relationship between bulge luminosity and central black-hole mass
in galaxies in the local universe has been used to claim that high mass
black holes are required for radio-loud QSOs (e.g. McLure and Dunlop
2001a,b, Dunlop et al 2001). Since the epoch of initial formation 
of the black hole and the `bulge' population are both unknown, as well as 
their subsequent change with time, this too should be treated with
caution. In addition, there is evidence that there is a continuum of radio
power in lower redshift QSOs (Lacy et al 2001), as well as systematic 
changes in radio
morphology with redshift (Neff and Hutchings 1990), so a simple dichotomy 
may not apply at high redshift either. Perhaps a more direct indication of
black hole mass is the broad-line width, but here too we need to understand
the kinematics of BLR, as well as its cosmic evolution.

   The irregular morphology of 4 of the five objects suggests that
the host galaxies may be in the process of heirarchical formation, or other 
tidal events, as well as associated with very active star-formation. This is 
similar to the conclusions of other investigations of z$\sim$2 QSOs.
The HST data have revealed structure of subarcsecond size, within 1-2
arcsec of the QSO nucleus that support and add to this scenario. These
results reinforce the caveats noted above.

   To the extent that the central black hole mass is related to the
initial stellar population in a galaxy (which in the present day universe
is the spheroidal, or non-disk, part of all galaxies), we may expect
the entire QSO host galaxy luminosity to be related to the central black
hole mass, at these redshifts where galaxies are very young, regardless 
of whether the morphology is strictly spheroidal. However, the 
proportionality factor of stellar luminosity to central mass may well change 
as the galaxy evolves (and undergoes merging) over its lifetime. We expect 
the spheroidal population luminosity to decline with time, with possible
boosts by major mergers, while the black hole mass can only increase - 
perhaps by a large factor.

   One way to check this independently is the widths of the nuclear 
broad emission lines, as discussed by McLure and Dunlop (2001a), 
among others. In our sample, there are 3 with published spectra, and 
these are not of very good quality for line width measures. The 
spectrum of 2244-010 (Foltz et al 1989) shows it to have
significant shortward absorptions to the broad lines, but we estimate the
width of the unabsorbed emission to have FWHM of 7000 km.s$^{-1}$.
We measure profiles in 0033+098 (Steidel and Sargent 1991) and 0820+296 
(Maoz et al 1993) to have FWHM 5400 and 3900
km.s$^{-1}$ respectively. In this small sample we see no correlation
with the host galaxy luminosity, or radio luminosity.

   The distribution of faint blue galaxies near the QSOs is similar
for all, with the exception of 0225-014, which has a much higher fraction
of companions within 10 arcsec (Table 2 gives the galaxies within the
entire $\sim$37 arcsec field of the PC). 
As a rigorous investigation of the distribution and detection 
criteria for companions seems unprofitable, we have not pursued these
statistics further. 

   Further progress awaits a large and uniform investigation of high
redshift QSO hosts, with high sensitivity as well as spatial resolution.
New instrumentation on HST as well as 8m class ground-based telescopes,
offer excellent opportunities for such studies.

\clearpage

\centerline{References}

Crampton D., Schade D., Cowley A.P., 1995, AJ, 90, 987

Dumont P.J., Redding D.C., Love, S., Boden A., Hanisch R.J., Mo J., 2001 
(in preparation)

Dunlop J.S., McLure R.J., Kukula M.J., Baum S., O'Dea C.P., Hughes D.H.,
2001 MNRAS (in press: astro-ph 0108397)

Foltz C.B., Chaffee F.H., Hewett P.C., Weymann R.J., Anderson S.F., 
MacAlpine G.M., 1989, AJ, 98, 1959

Heckman T.M., Lehnert M.D., van Breugel W., Miley G.K., 1992, ApJ, 370, 78

Hewitt A., and Burbidge G., 1993, ApJS, 87, 451

Hutchings J.B., 1995a, AJ, 109, 928

Hutchings J.B., 1995b AJ, 110, 994

Hutchings J.B., 2001a, astro-ph 0107157

Hutchings J.B., 2001b, IAU colloquium 184 (in press)

Krist, J. 1995, ASP Conf Ser 77 (ADASS IV), R. A. Shaw, H. E. Payne, \& J. J.
E. Hayes, eds., p. 349

Kukula M.J. et al 2001 MNRAS (in press: astro-ph 0010007)

Lacy M., Laurent-Muehmeisen S.A., Ridgway S.E., Becker R.H., White R.L.,
2001

Lehnert M.D., Heckman T.M., Chambers K.C., Miley G.K., 1992, ApJ, 393, 68

Maoz D., Bahcall J.N., Schneider D.P. et al 1993, ApJ, 409, 28

McLure R.J., Dunlop J.S., Kukula M.J., Baum S.A., O'Dea C.P., Hughes D.H.,
1999, MNRAS, 308, 377

McLure R.J., and Dunlop J.S., 2001a, MNRAS, 327, 199

McLure R.J., and Dunlop J.S., 2001b, astro-ph 0108417

Neff S.G., and Hutchings J.B., 1990, AJ, 100, 1441

Pauliny-Toth I.I.K., Kellerman K.I., Davis M.M., Fomalont E.B., and
Shaffer D.B., 1972, AJ, 77, 265 

Ridgway S.E., Heckman T.M., Calzetti D., Lehnert M., 2001, ApJ, 550, 122

Steidel C.C. and Sargent W.L.W., 1991, AJ, 102, 1610

\newpage 

\centerline{Captions to figures}

1. PSF models with different offsets of HST secondary mirror. The
offset difference is 2 mm. Note the significant changes in the brightness
of the inner ring of spots and the components on the main diffraction
spikes. A grid of these models was used to match the structure seen in
the QSO images.

2. PSF-star images to illustrate subtraction. The PSF model and star
image are shown in the top row, matched in pixel sampling, and the 
subtracted image is below. The images
are 3.2 arcsec on a side. The text notes that this PSF subtraction is not
as good as for the QSO images.

3. Images of the program QSOs, 2.8 arcsec on a side. These are 
after optimal PSF subtraction, as described in the text, and are 
combined from the two filter images. The images have
been smoothed with a 0.037 arcsec gaussian to reduce pixel noise. The 
saturated inner cores in most of them are dominated by PSF artifacts
but the core of the faint QSO 1338+277 is resolved. Note the asymmetric
structures, the knot to the left of 0033+098
and the `jet' of 1338+277. Faint traces of the diagonal diffraction spikes
are seen in some images. North is 126$^o$ clockwise of up for 0033+098;
17$^o$ anticlockwise for 0225-014; 85$^o$ clockwise for 0820+296; 
100$^o$ clockwise for 1338+277; 30$^o$ clockwise for 2244-010.

4. Luminosity profiles of sample QSOs. Top left shows full QSO image
showing it is resolved compared with the PSF. The other profiles are
after PSF subtraction. Lowest levels plotted are close to 27 mag per 
square arcsec. The excess light within r$^{1/4}\sim$0.65 is
not significant and ignored in estimating the host galaxy fluxes. The
dotted lines are the spheroidal models used to calculate the `bulge'
fluxes. Excess light above these lines at larger radii is fitted with
an exponential to calculate the detected `disk' components.


\begin{deluxetable}{llllcc}
\tablecaption{Observations}
\tablehead{\colhead{Name} &\colhead{Mag\tablenotemark{1}} &\colhead{z}
&\colhead{Date} &\colhead{F606W\tablenotemark{2}} 
&\colhead{F702W\tablenotemark{2}} }
\startdata
0033+098\tablenotemark{3} &17.5 &1.91 &Nov 22 1998 &4650 &5200\nl
0225$-$014\tablenotemark{3} &18.2 &2.04 &Nov 12 1998 &5000 &5400\nl
0820+296\tablenotemark{3} &18.5 &2.37  &Feb 20 1999 &5300 &5400\nl
1338+277 &20.0 &2.28 &Feb 14 1999 &5300 &5400\nl
2244$-$010 &18.0 &2.03 &May 15 1999 &5000 &5400\nl
Feige 23 &11.1 &0    &Oct 24 1998 &$\sim$900 &$\sim$900\nl
\enddata
\tablenotetext{1}{Catalogue V magnitude}
\tablenotetext{2}{Exposure time in seconds}
\tablenotetext{3}{Radio-loud QSO}
\end{deluxetable}

\begin{deluxetable}{lccccc}
\tablecaption{Summary of results}
\tablehead{&\colhead{0033+098} &\colhead{0225-014} &\colhead{0820+296}
&\colhead{1338+277} &\colhead{2244-010} }
\startdata
{\bf~~~F606W} \nl
Total &18.9 &18.7 &19.1 &21.5 &17.6\nl
Bulge &20.1 &21.3 &21.3 &23.1 &20.6\nl
Disk &22.9 &23.9 &24.3 &24.6 &22.9\nl
Bulge/Total &0.36 &0.09 &0.14 &0.23 &0.06\nl
\nl
{\bf~~~F702W} \nl
Total &18.6 &18.6 &19.2 &21.6 &17.5\nl
Bulge &20.2 &21.3 &21.1 &23.5 &20.0\nl
Bulge/Total &0.24 &0.08 &0.17 &0.17 &0.10\nl
\nl
QSO R-I &0.3 &0.1 &-0.1 &-0.1 &0.1 \nl
Bulge R-I &0.1 &0.0 &0.2 &-0.4 &0.6 \nl
Bulge M$_R$\tablenotemark{a} &-24.7 &-23.8 &-24.1 &-22.3 &-24.5\nl
Disk M$_R$ &-21.9 &-21.2 &-21.1 &-20.8 &-22.2\nl
\nl
Comp\#\tablenotemark{b} &3/10 &6/15 &3/11 &4/11 &2/12\nl
Comp dist (")\tablenotemark{c} &15$\pm$4 &13$\pm$5 &14$\pm$5 
&12$\pm$4 &15$\pm$5\nl
Comp mag\tablenotemark{d} &24.6$\pm$1.0 &25.0$\pm$0.8 
&24.0$\pm$1.2 &24.7$\pm$0.4 &25.1$\pm$1.0 \nl
Comp R-I\tablenotemark{e} &0.24$\pm$0.58 &-0.1$\pm$0.43 &0.2$\pm$0.3
&0.0$\pm$0.34 &0.25$\pm$0.33 \nl
Comp mean M$_R$ &-20.2 &-20.1 &-21.4 &-20.7 &-20.0\nl 

\enddata
\tablenotetext{a}{M values with no k-correction}
\tablenotetext{b}{Galaxies in both filters, within 10" and over whole PC field}
\tablenotetext{c}{Average offset from QSO in arcsec}
\tablenotetext{d}{Average R mag of galaxies}
\tablenotetext{e}{Average R-I of galaxies $<$1.0 - i.e. w/o foreground}

\end{deluxetable}

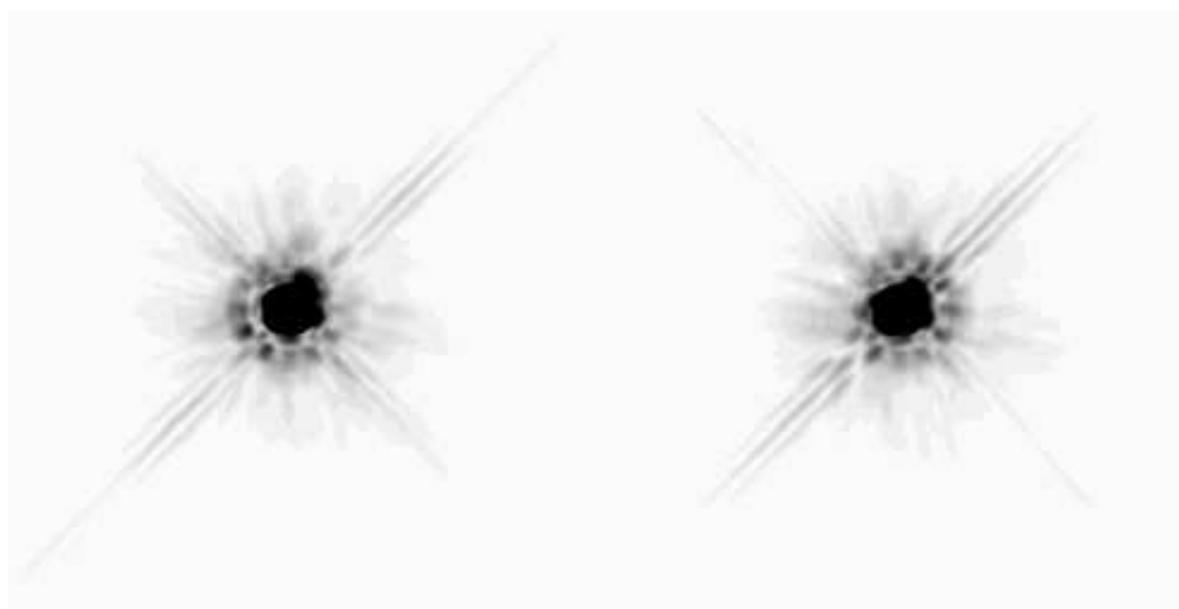
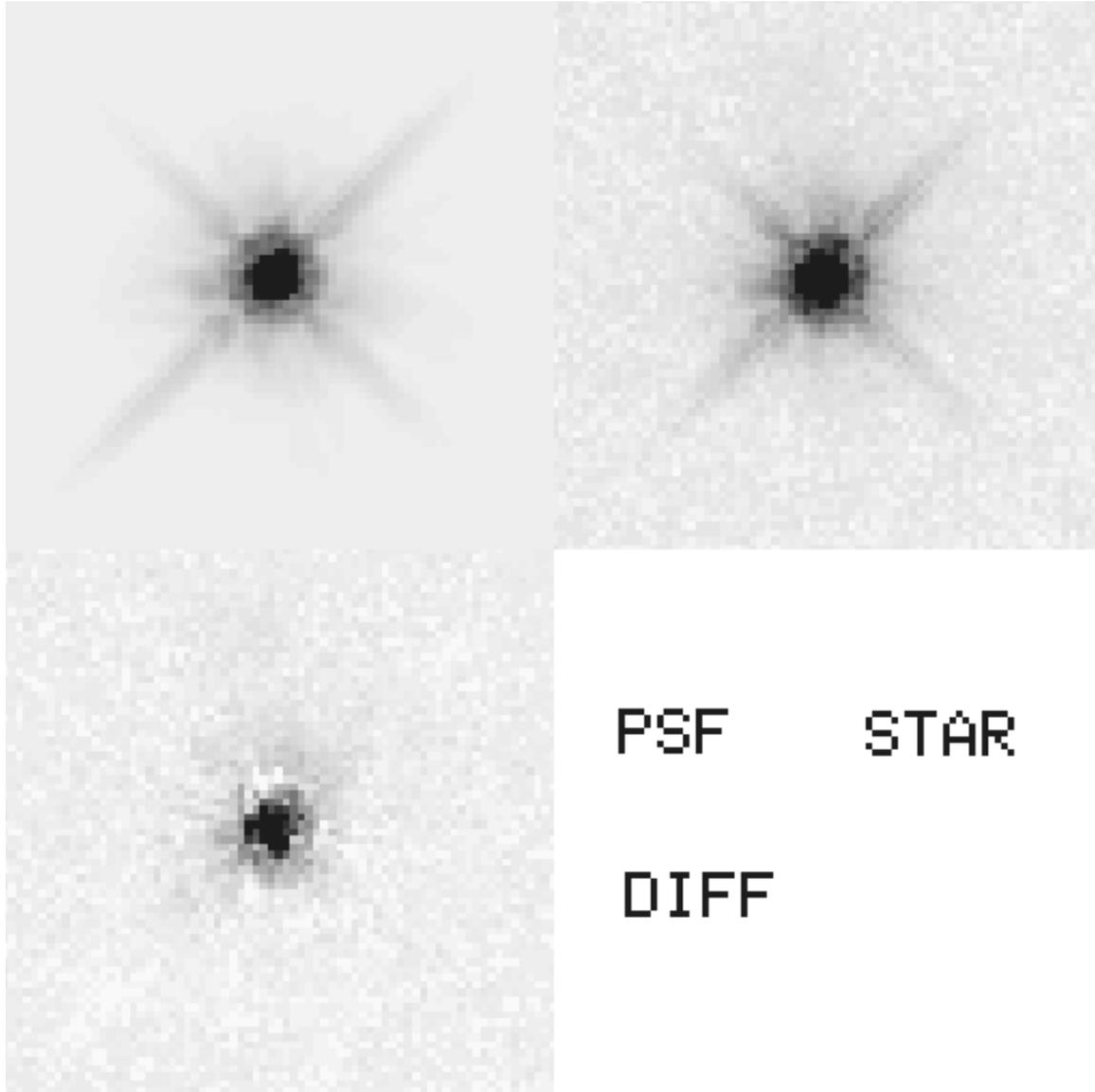
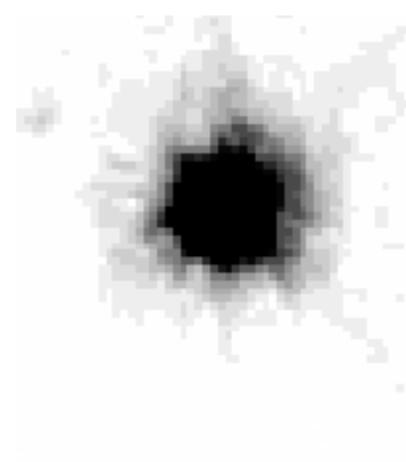
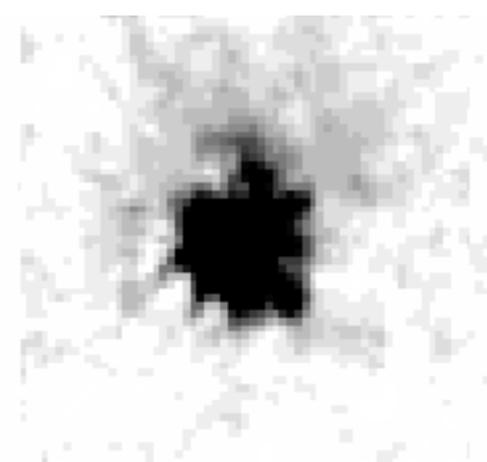
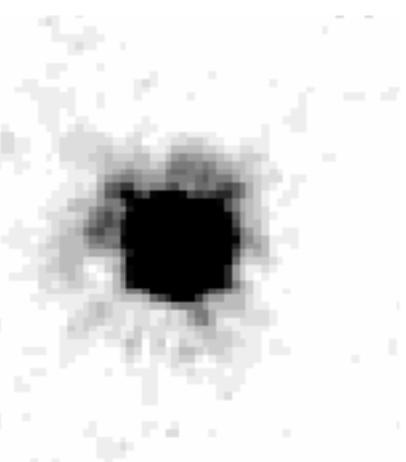
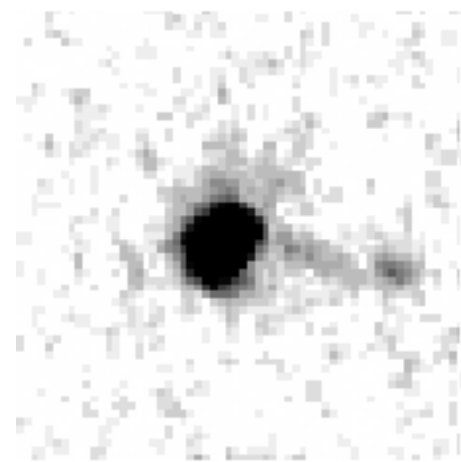
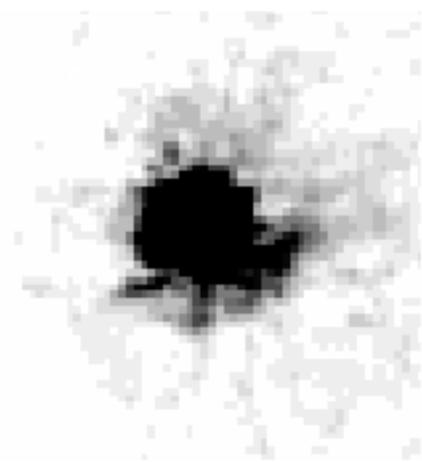
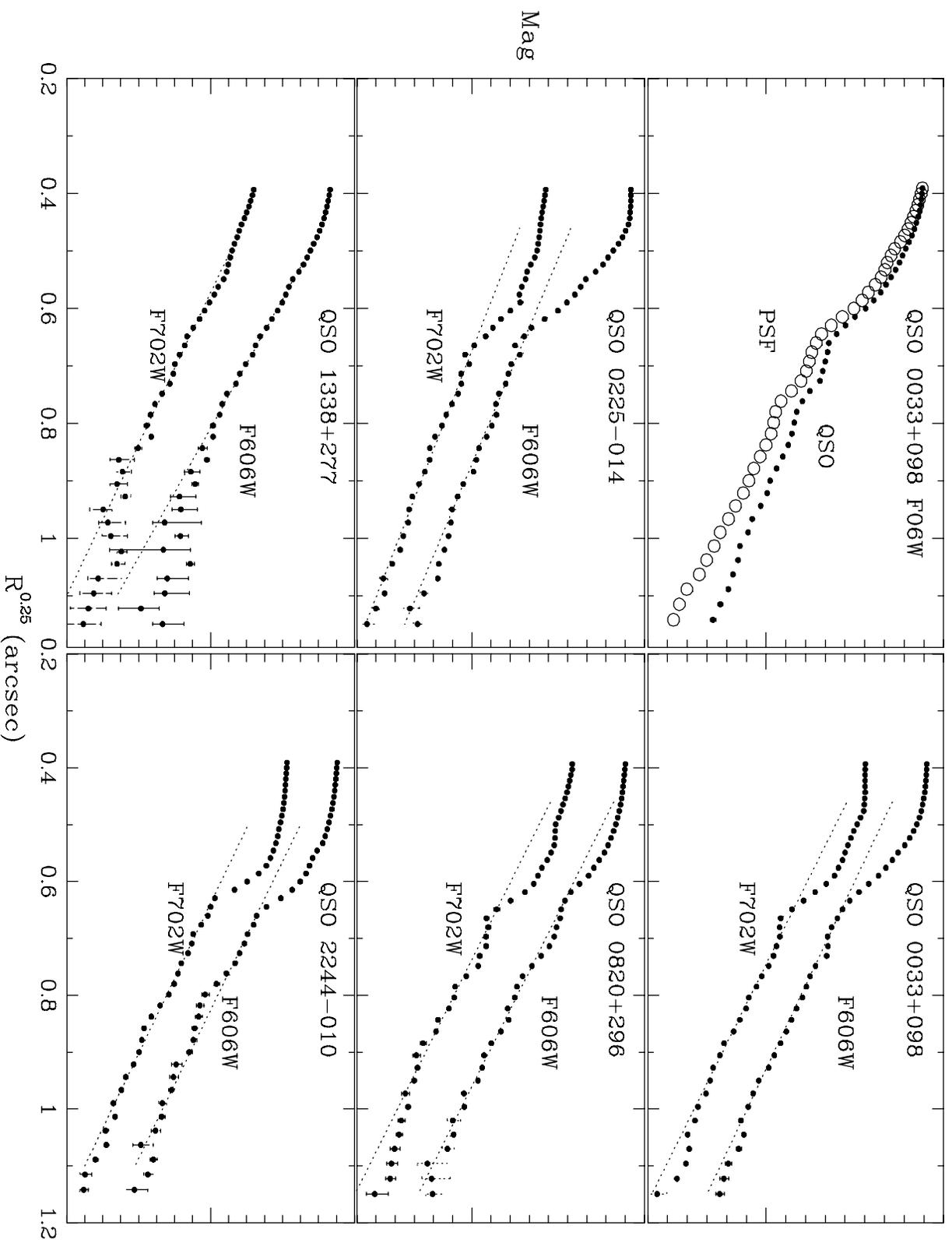
\end{document}